\documentclass[twocolumn,pre,showpacs,preprintnumbers,amsmath,amssymb,superscriptaddress]{revtex4}

\usepackage{graphicx}% Include figure files
\usepackage{dcolumn}% Align table columns on decimal point
\usepackage{bm}% bold math
\usepackage{longtable}% bold math

\begin{document}

\preprint{-}

\title{Scattering lengths and universality in superdiffusive L\'evy materials}

\author{}
\affiliation{}

\author{Raffaella Burioni}
\affiliation{Dipartimento di Fisica, Universit\`a degli Studi di
Parma, viale G.P.Usberti 7/A, 43100 Parma, Italy}
\affiliation{INFN, Gruppo Collegato di
Parma, viale G.P. Usberti 7/A, 43100 Parma, Italy}
\author{Serena di Santo}
\affiliation{Dipartimento di Fisica, Universit\`a degli Studi di
Parma, viale G.P.Usberti 7/A, 43100 Parma, Italy}
\author{Stefano Lepri}
\affiliation{CNR-Istituto dei Sistemi Complessi, 
via Madonna del Piano 10, I-50019 Sesto Fiorentino, Italy}
\author{Alessandro Vezzani}
\affiliation{
Centro S3, CNR-Istituto di Nanoscienze, Via Campi 213A, 41125 Modena Italy}
\affiliation{Dipartimento di Fisica, Universit\`a degli Studi di
Parma, viale G.P.Usberti 7/A, 43100 Parma, Italy}

\date{\today}

\begin{abstract}
We study the effects of scattering lengths on L\'evy walks in quenched one-dimensional
random and fractal quasi-lattices, with scatterers
spaced  according to a long-tailed distribution.  By analyzing the scaling
properties of the random-walk probability distribution, 
we show that the effect of the varying scattering length can be
reabsorbed in the multiplicative coefficient of the scaling length. This leads to a superscaling
behavior, where the dynamical exponents and also the scaling functions
do not depend on the value of the scattering length. Within the scaling framework, we obtain an exact expression for the multiplicative coefficient as a function of the scattering length both in the annealed and in the quenched random and fractal cases. Our analytic results are compared with numerical simulations, with  excellent agreement, and are supposed to hold also in higher dimensions.
\end{abstract}

\pacs{5.40.fb 02.50.Ey 05.60.k} \maketitle

\section{\label{sec:intro}Introduction}

Diffusion in heterogeneous and porous materials \cite{havlin}, composed of two or more types
of regions with very different diffusion properties, can be described  as a
sequence of independent scattering events occurring in the  hard-scattering part
of the material, followed by long jumps performed  at almost constant velocity
in non-scattering regions. The  single-particle dynamics thus amount to a
random walk, where each step length $l$ is a random variable with a
given  probability distribution $\lambda(l)$. As it is well known, if $l$ has  a
finite variance, the ensuing process follows the standard laws of Brownian
motion. If instead the material is very heterogeneous on all scales, the step
length distribution may become heavy-tailed and when   $\langle l^2\rangle$
diverges the diffusion can become  anomalous \cite{bouchaud,klages}. Whenever the
detailed structure of the  underlying scattering media can be ignored 
(\textit{annealed} disorder),  the dynamics is only ruled by the behavior of $\lambda(l)$ for
large $l$ and the model corresponds to a standard L\'evy walk \cite{shles1985,ann,klafterec}.  If
instead one takes into account that the steps are correlated by their 
mutual positions in the sample, the step length distribution represents 
a \textit{quenched} disorder. It is argued that quenched and annealed 
disorder differ in many respects \cite{Fogedby94,klafter,been2,buonsante}. 
  
A particularly interesting application is multiple scattering of light in
disordered media, which can often be described as a random walk process,
analogous to the Brownian motion of massive particles \cite{Sheng}.  In that
case, if scattering elements are homogeneously distributed in space, one
usually observes diffusive transport. However, when the local
density of scattering elements is strongly inhomogeneous, the optical transport
properties of a material can change dramatically and lead to superdiffusion,  as
exemplified for instance by photon diffusion through clouds \cite{davis}. More
recently, artificial materials have been assembled in the lab that allowed for
an unprecedented experimental demonstration of light superdiffusion \cite{Levy,bertolotti}. 
Interestingly, in experiments  a change in the optical density of the 
diffusive media allows to tune the so called scattering length (or scattering mean 
free path) \cite{barthelemy,bertolotti}, which is a measure of the probability of
experiencing a scattering event \cite{Sheng}. Also Monte Carlo simulations 
\cite{barthelemy} indicate that this is an important parameter, that may significantly
affect the observability of asymptotic scaling regimes. It is thus important to asses 
the role of the scattering lengths on scaling properties  of relevant observables, 
starting with probability distributions, to interpret correctly experiments and
simulations in inhomogeneous scattering media. While in pure diffusive
processes a varying scattering length can be simply encoded in a trivial change
of the variance of the Gaussian distributions, superdiffusive process have not
been investigated in details. 

Many relevant features of experiments on scattering in inhomogeneous media
can be described 
as a random walk in a quenched, long-range correlated environment. 
The simplest case consists of a free particle moving 
through a one-dimensional array of barriers whose spacing is power-law distributed 
\cite{beenakker,falceto,Levyrandom,Cantor1,Cantor2}. 
To address the question discussed above, in this work we extend such models to the
case in which the transmittance through each barrier can be tuned to be 
different from $1/2$. This models a situation in which the 
velocity in the scattering media is not fully randomized at each collision.  
Changing the barrier transmittance is thus akin to changing the scattering length
in the diffusive portion of the material. 

In the following, we will show that
the probability distribution for the random walker to be a time $t$ in the point $r$
exhibits a scaling form and that the effect of varying the transmittance can be
reabsorbed in the multiplicative coefficient of the scaling length of the 
process. Numerical simulations confirm this superscaling behavior and evidence that not only the dynamical exponents are universal, as expected from the general scaling framework, but also the scaling functions are unchanged. Notice that the latter are in general not Gaussian in these systems.  
Moreover, we obtain an analytic expression for the multiplicative coefficient of the scaling length 
as a function of the transmittance both in the annealed and in the quenched models;
in the annealed case the analytic form  is calculated by solving directly 
the master equation of the process; in the quenched models it can be derived, within the scaling hypothesis, by considering an extended Einstein relation between the stationary conduction and the (super)diffusion.
Interestingly such coefficient turns out to be different in the quenche and in the annealed models and, in the latter case, it is independent of the transmittance when superdiffusion is present. This evidences another relevant difference between the two approaches.
Finally, we apply the same scaling picture,  as a function of the transmittance,
to the time resolved transmitted intensity through a finite sample of length $L$ 
and to the total transmission,  which are the quantities usually measured in experiments.

The paper is organized as follows: in the next section we define the random and fractal quasi-lattices models with arbitrary transmittance, introducing the scaling picture and the superuniversal behavior. 
In Section \ref{sec:annealed} we calculate the exact expression, as a function of the transmittance, of the multiplicative coefficient for the scaling length in the annealed case. In Section
\ref{sec:quenched} we turn to the quenched random and fractal quasi-lattices and we obtain an exact expression for the coefficients in these cases. In Section \ref{sec:numerics} we analyze the superscaling framework for random structures with L\'evy parameter $0< \alpha<1$,  a case with non Gaussian scaling functions. Then, we focus on random structures with $1<\alpha<2$, where the Gaussian scaling function obeys  superscaling, but a subleading term determines the moments of the distribution.
We evidence that also this subleading term can be studied 
in terms of a change in the multiplicative constant for the scaling length. 
We finally verify the superscaling framework in the deterministic case with $0<\alpha<1$ where 
the scaling function presents log-periodic oscillations  \cite{Cantor1,Cantor2}. 
In Section \ref{sec:timeres} we apply our superscaling approach to a different physical quantity, 
the time resolved transmission intensity, which is relevant for a realistic experimental setup 
in inhomogeneous optical media. In the last Section we present 
our conclusions.

\section{\label{sec:models }L\'evy Walks on random and fractal quasi-lattices}

\begin{figure}
\includegraphics[width=\columnwidth]{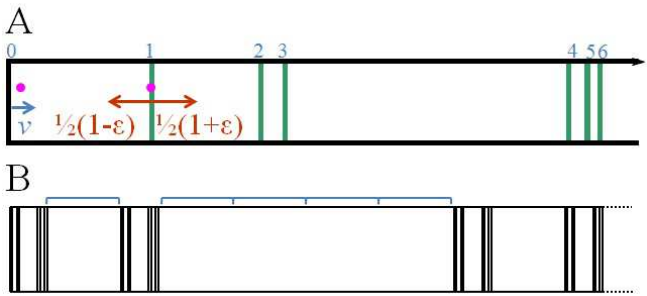}
\caption{(a) Example of a random structure, where scatterers are distributed according
equation \eqref{p_r}.  A particle with velocity $v$ is transmitted with
probability $(1+\varepsilon)/2$ and reflected with probability
$(1-\varepsilon)/2$; (b) A deterministic Cantor like L\'evy quasi-lattice,
characterized by $n_r=2$ and $n_u=4$.}
\label{model}
\end{figure}

We consider a continuous time random walk on one-dimensional structures 
(see Fig. \ref{model}), where point-like scatterers are spaced according 
to a L\'evy distribution. The walker moves
with constant velocity $v$ in between each two consecutive 
scatterers.  When it arrives at a scatterer, it can reverse its direction
of motion ($v \to -v$) with probability  $(1-\varepsilon)/2$ (with $-1<\varepsilon<1$).  
Clearly,  $|v|$ is conserved  during
the evolution and we can set $|v|=1$ without loss of generality. 

We investigate two types of structures. The first is random and (upper panel of
Fig. \ref{model}) 
the probability for two consecutive scatterers, labeled by
the indexes $j$  and $j+1$, to be
at distance $r$ is \cite{klafter,beenakker,Levyrandom}
\begin{equation}
  \label{p_r}
  \lambda(r)\equiv \frac{\alpha r_0^\alpha}{r^{\alpha+1}}, \quad r\in [r_0,\infty),
\end{equation}
where $\alpha>0$ and $r_0$ is a cutoff fixing the scale length of the system. 
The second type is a class of deterministic quasi-lattices (lower panel of
Fig. \ref{model}), built by placing the scatterers on generalized Cantor 
sets \cite{Cantor1,Cantor2}. Each set, and the ensuing step length distribution,
is defined by the two parameters $n_u$ and $n_r$ used in its recursive construction. 
The former represents the growth of 
the longest step when the structures is increased by a generation, so that the longest
step in a structure of generation $G$ is proportional to $n_u^G$; $n_r$ is the number 
of copies of generation $G-1$ that form the generation $G$, so that the total number
of scatterers in the generation $G$ is proportional to $n_r^{G}$ (see Ref.\cite{Cantor1}
for details). For this second type of structures,
the role of the exponent $\alpha$ of the random case 
is played by $\alpha = \log n_r /  \log n_u$ \cite{Cantor2}.

In the random case, we will average over different realizations of the 
structure and  we will consider  averages  taken  over processes starting from
scattering sites. For  quenched L\'evy processes, it is known that different
averaging procedures can lead to  different behaviors. Moreover, properties
arising from averages taken over processes starting in any  point  are different
\cite{klafter,Cantor2,Levyrandom,buonsante}. In the deterministic case, we consider 
averages performed over random trajectories starting from a given point,
e.g the origin $0$ evidenced 
in Fig. \ref{model}.

The main quantity we are interested in is the probability for a 
walker to be at time $t$ a distance $r$ from the starting point, which we 
denote by $P_{\alpha,\varepsilon}(r,t)$ to emphasize the dependence 
on the two basic parameters of this class of models, $\alpha$ and $\varepsilon$. 
The process on a L\'evy structure with given $\alpha$ can be described by introducing
a scaling function $f_{\alpha}(x)$ and a scaling  
length $\ell_\varepsilon(t)$ growing with time \cite{Levyrandom}. The scaling form
for $P_{\alpha,\varepsilon}(r,t)$ can be written as:
\begin{equation}
P_{\alpha,\varepsilon}(r,t)=\ell_\varepsilon^{-1}(t)f_{\alpha}(r/\ell_\varepsilon(t)) + h_{\alpha,\varepsilon}(r,t)
\label{sal}
\end{equation}
where $h_{\alpha,\varepsilon}(r,t)$ is a function that vanishes in probability for large 
times: 
\begin{equation}
\lim_{t\to\infty}\int_0^{v t} |P_{\alpha,\varepsilon}(r,t)-\ell_\varepsilon^{-1}(t)f_{\alpha}(r/\ell_\varepsilon(t))|dr=0
\label{sal2}
\end{equation}
In \cite{Levyrandom}, the scaling form \eqref{sal},\eqref{sal2}, has been analyzed in details and it has been shown to hold for $\varepsilon=0$. Moreover, the growth of $\ell_0(t)$ has been shown to follow the asymptotic law $\ell_0(t)\sim t^{1/(1+\alpha)}$ for $0<\alpha<1$
and $\ell_0(t)\sim t^{1/2}$ for $\alpha>1$. 
The presence of the subleading function $h_{\alpha,0}(r,t)$ and of  long tails in $f_{\alpha}(x)$,  induce a nontrivial scaling of higher-order momenta, $\langle r^p(t)\rangle \not\simeq \ell^p(t)$, leading to strongly anomalous diffusion \cite{castiglione}. 

Here we will evidence that, within a superscaling framework,
the dynamical  exponents and the scaling functions are independent of the transmittance $\varepsilon$.
In particular, ${f}_\alpha(\cdot)$ is independent of  $\varepsilon$ and the scaling length reads:
\begin{equation}
\ell_\varepsilon(t) \simeq
\begin{cases}
A_{\varepsilon} t^{\frac{1}{1+\alpha}} & \mathrm{if}\ 0<\alpha<1 \\
A_{\varepsilon} t^{1 \over 2} & \mathrm{if}\ 1 \leq \alpha  
\label{ellcom}
\end{cases}
\end{equation}
Notice that $\ell_\varepsilon(t)$ is defined up to an arbitrary multiplicative constant, that in numerical simulation will be fixed equal to one for $\varepsilon=0$. 
Within the hypothesis that  ${f}_\alpha(\cdot)$ is independent of $\varepsilon$, in the next sections we will obtain an analytic expression for $A_\varepsilon$ both in the annealed and in the quenched case.

\section{\label{sec:annealed} Transmittance and Annealed L\'evy Walks}

Before discussing the effect of the transmittance $\varepsilon$ in quenched systems, 
we consider the annealed case, where the length of the ballistic stretches is chosen randomly
from the distribution \eqref{p_r} independently at each scattering event. In this case  the topology of
the system is not taken into account and only the distribution of the steps influences
the dynamics. In this simpler situation, an analytical approach is feasible and the behavior of the scaling length
as a function of the transmittance can be determined. Let us introduce
$P^+(r,t)$ and $P^-(r,t)$ as the probabilities of being in $r$ at time $t$ arriving 
respectively  from the left and from the right. In the persistent random walk approach, recalling that $|v|=1$, we 
can write
\begin{eqnarray}
 & & P^+(r,t) =  \int_{r_0}^\infty \left[ \frac {1+\varepsilon} 2   P^+(r-r',t-r')+ \right. \nonumber \\ 
& & + \left. \frac {1-\varepsilon} 2   P^-(r-r',t-r') \right] \lambda(r') dr' + \frac 1 2 \delta(r,0) \delta (t,0)\nonumber \\
 & & P^-(r,t) =  \int_{r_0}^\infty \left[ \frac {1+\varepsilon} 2   P^-(r+r',t-r')+ \right.  \nonumber \\ 
& & + \left. \frac {1-\varepsilon} 2   P^+(r+r',t-r') \right] \lambda(r') dr' + \frac 1 2 \delta(r,0) \delta (t,0).\nonumber\\
\label{Qmaster}
\end{eqnarray}
By applying a Fourier transform in space and time in \eqref{Qmaster}, we obtain an expression 
for $\tilde P(k,\omega)$, that is the transformed of $P_{\alpha,\varepsilon}(r,t)=P^+(r,t)+P^-(r,t)$.
In particular in the asymptotic regime for large space and times, i.e. for $\omega$ and $k$ going to zero, we have
the following estimates:
\begin{equation}
P_{\alpha,\varepsilon}(k,\omega)=
\begin{cases}
\left(C_1 \omega^\alpha + C_2 k^\alpha\right)^{-1} & \mathrm{if}\ 0<\alpha<1 \\
 \left (C_3 \omega + C_4 k^\alpha \right )^{-1} & \mathrm{if}\ 1 < \alpha < 2\\
\left (C_5 \omega + C_6 \frac{1+\varepsilon C_7}{1-\varepsilon} k^2 \right)^{-1}  & \mathrm{if}\ 2 < \alpha
\end{cases}
\label{Qan}
\end{equation}
where the complex constants $C_1 \dots C_7$ depend on the distribution $\lambda(r)$ (i.e. on $\alpha$) but are independent of $\varepsilon$. The asymptotic scaling form of $P_{\alpha,\varepsilon}(r,t)$ can then be obtained by inverting the Fourier transform in \eqref{Qan}. In particular we have:
\begin{equation}
P_{\alpha,\varepsilon}(r,t)\simeq \ell_\varepsilon(t) ^{-1} \tilde{f}_\alpha(r/\ell_\varepsilon(t))
\label{Panscal}
\end{equation}
with
\begin{equation}
\ell_\varepsilon(t)\simeq
\begin{cases}
t  & \mathrm{if}\ 0<\alpha<1 \\
t^{1/\alpha} & \mathrm{if}\ 1 < \alpha < 2\\
\left (\frac{1+\varepsilon C_7}{1-\varepsilon}\right)^{1/2} t^{1/2}  & \mathrm{if}\ 2 < \alpha
\end{cases}
\label{ell_an}
\end{equation}
where the function $\tilde{f}_\alpha(\cdot)$ and proportionality constants in \eqref{ell_an} 
are independent of $\varepsilon$. 

\begin{figure}
\includegraphics[width=\columnwidth]{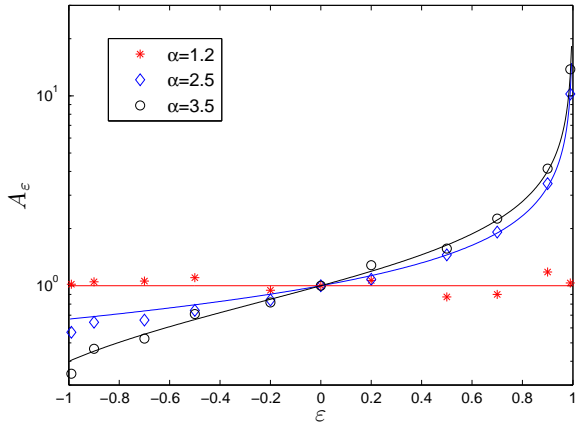}
\caption{ Value of the generalized diffusivity in \eqref{ell_an} as a function of $\varepsilon$. Symbols represent numerical simulations, while continuos lines are the analytical calculations.}
\label{A_epsilon_an}
\end{figure}

As expected, the dynamical exponents in equation \eqref{ell_an} 
coincides with the well known results for the annealed L\'evy  walks with $\varepsilon=0$, established in \cite{klafterec}. 
In addition, equations 
(\ref{Panscal},\ref{ell_an}) show that both the dynamical exponents 
and the scaling function $\tilde{f}_\alpha(\cdot)$ are independent of $\varepsilon$, 
even in the non trivial case of anomalous diffusion i.e. $\alpha<2$. Moreover, 
quite surprisingly for $\alpha<2$, i.e. in the ballistic and superdiffusive cases, even the coefficient of the scaling length 
does not depend on the transmittance and therefore the whole asymptotic regime is 
independent of the value of $\varepsilon$. On the other hand, for $\alpha>2$ the same coefficient depends in a non trivial way both on $\varepsilon$ and on the step length distribution $\lambda(r)$. Indeed, $C_7 =(2\langle r \rangle^2 / \langle r^2 \rangle-1)$ where $\langle r^k \rangle =\int r^k \lambda (r) dr$. In particular, for the step lengh distribution considered in \eqref{p_r},
$C_7=(\alpha^2-2\alpha-1)/(\alpha-1)^2$. Notice that the diffusivity diverges for $\varepsilon=1$ (perfect transmission), while for $\varepsilon=-1$ it vanishes only for $C_7=1$, hence only when the step lengths do not fluctuate. Indeed, step length fluctuations induce diffusion even in the case of total reflection.
Fig. \ref{A_epsilon_an} compares the analytical prediction for the coefficient in Equations \eqref{ell_an} with numerical simulations at different value of $\alpha$ showing an excellent agreement.

\section{\label{sec:quenched}Transmittance and quenched  L\'evy Walks}

\begin{figure}
\includegraphics[width=\columnwidth]{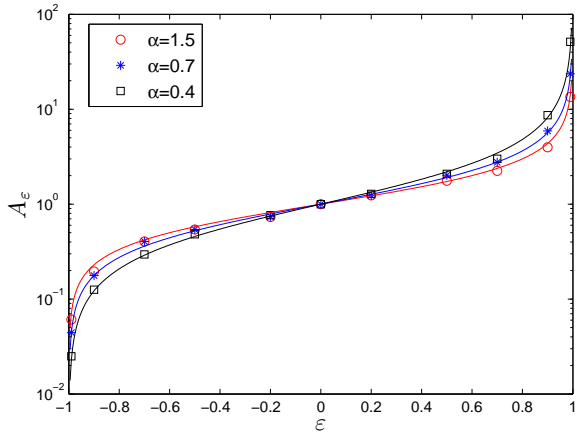}
\caption{ Value of the coefficient $A_{\varepsilon }$ as a function of $\varepsilon$. Symbols represent numerical simulations, while continuos lines are the analytical calculations of formula \eqref{Aepsilon}.}
\label{A_epsilon}
\end{figure}

Let us now turn to the quenched case. Within the scaling framework described by equations (\ref{ellcom},\ref{sal}), we first derive an analytic expression for the coefficients 
$A_{\varepsilon}$. We can exploit the fluctuation-response relation connecting 
$P_{\alpha,\varepsilon}(r,t)$ to $C_{\alpha,\varepsilon}(L)$ i.e. the stationary conductivity of a 
system of size $L$. In particular according to \cite{Cantor1,cates} we have:
\begin{equation}
C_{\alpha,\varepsilon}(L)^{-1}=\lim_{\omega\to 0} \int {\rm e}^{i\omega t} (P_{\alpha,\varepsilon}(L,t)- 
P_{\alpha,\varepsilon}(0,t)) dt
\label{einst}
\end{equation}
Plugging the scaling form \eqref{sal} of $P_{\alpha,\varepsilon}(r,t)$ into equation \eqref{einst} and imposing that the scaling function depends on the transmittance $\varepsilon$ only through the constant $A_{\varepsilon}$, as in \eqref{ellcom}, we obtain:
\begin{equation}
C_{\alpha,\varepsilon}(L) \simeq 
\begin{cases}
A_{\varepsilon}^{1+\alpha}\, L^{-\alpha} & \mathrm{if}\ 0<\alpha<1 \\
A_{\varepsilon}^{2}\, L^{-1} & \mathrm{if}\ 1 \leq \alpha  
\label{cond3}
\end{cases}
\end{equation}
where the proportionality constant is independent of $\varepsilon$. Equation \eqref{cond3} extends the Einstein relation for anomalous conductivity in \cite{Cantor1,cates}, taking into account the transmittance $\varepsilon$. The conductivity can be evaluated directly by studying the stationary current in a system of size $L$ and fixed boundary conditions. The (stationary) master equation reads:
\begin{eqnarray}
 & & P_{\alpha,\varepsilon}^+(r_k) =   \frac {1+\varepsilon} 2   P^+(r_{k-1})+  
\frac {1-\varepsilon} 2   P^-(r_{k-1})  \nonumber \\
 & & P_{\alpha,\varepsilon}^-(r_k) =  \frac {1+\varepsilon} 2   P^-(r_{k+1})
+ \frac {1-\varepsilon} 2   P^+(r_{k+1}) 
\label{Qmaster_st}
\end{eqnarray}
where $P_{\alpha,\varepsilon}^+(r_k)$ and $P_{\alpha,\varepsilon}^-(r_k)$ represent the stationary probabilities of being at the 
scattering site $r_k$ arriving from the left and from the right, respectively.
The solution of equation \eqref{Qmaster_st} is 
\begin{eqnarray}
 & & P_{\alpha,\varepsilon}^+(r_k) =  a k -  \frac a {1-\varepsilon} +c \nonumber \\
 & & P_{\alpha,\varepsilon}^-(r_k) =  a k +  \frac a {1-\varepsilon} +c
\label{Qmaster_stsol}
\end{eqnarray}
where $a$ and $c$ are arbitrary constants. Imposing  the particle density at the 
borders $P_{\alpha,\varepsilon}(0)=P_{\alpha,\varepsilon}^+(0)+P_{\alpha,\varepsilon}^+(0)=1$ and $P_{\alpha,\varepsilon}(L)=P_{\alpha,\varepsilon}^+(L)+P_{\alpha,\varepsilon}^+(L)=0$ we get $c=1/2$ and $a=(2K)^{-1}$ where $K$ is the number of scatterers between $0$ and $L$. Since the conductivity equals the total current flowing in the system  we have
\begin{equation}
C_{\alpha,\varepsilon}(L)=P_{\alpha,\varepsilon}^+(r_{k+1})-P_{\alpha,\varepsilon}^-(r_k)= \frac {1+\varepsilon} {2K(1-\varepsilon)}
\label{einst2}
\end{equation}
and averaging over different disorder realizations according to \cite{beenakker} we have
\begin{equation}
C_{\alpha,\varepsilon}(L)= 
\begin{cases} 
D_\alpha \frac {1+\varepsilon} {L^\alpha(1-\varepsilon)}& \mathrm{if}\ 0<\alpha<1 \\
D_\alpha \frac {1+\varepsilon} {L(1-\varepsilon)}& \mathrm{if}\ 1<\alpha \\
\end{cases} 
\label{einst3}
\end{equation}
where the constant $D_\alpha$ is independent of $\varepsilon$. The results in equations \eqref{einst3} is consistent with  \eqref{ellcom} and \eqref{cond3}, with an analytical estimate for the multiplicative coefficients of the scaling lenghts:
\begin{equation}
A_{\varepsilon}= 
\begin{cases} 
\left(\frac {1+\varepsilon}{1-\varepsilon}\right)^{\frac 1 {1+\alpha}}& \mathrm{if}\ 0<\alpha<1 \\
\left(\frac {1+\varepsilon}{1-\varepsilon}\right)^{\frac 1 2}& \mathrm{if}\ 1<\alpha \\
\end{cases} 
\label{Aepsilon}
\end{equation}
In Fig. \ref{A_epsilon} we compare our analytical prediction \eqref{Aepsilon} with numerical simulations at different value of $\alpha$, with excellent agreement. Clearly $A_{\varepsilon}$ diverges for $\varepsilon\to 1$, i.e. perfect transmission and $A_{\varepsilon}$ vanishes for $\varepsilon\to -1$, i.e. total reflection. We remark that $A_{\varepsilon}$ behaves differently than in the annealed model both in the superdiffusive and in the normal case.

\section{\label{sec:numerics} Numerical evidences on quenched structures}

Let us now present a numerical analysis of the quenched cases, evidencing that our assumption holds, 
i. e. that the whole effect of a variation in 
$\varepsilon$ can be summarized in a variation of the coefficient 
$A_{\varepsilon}$, and therefore that the scaling functions are super-universal, 
i.e. $f_{\alpha}(x)$ in Eq. \eqref{sal} does not depend on 
$\varepsilon$. For standard diffusion this properties obviously holds:
indeed the  scaling functions are Gaussian and they can be characterized basically by their variance, 
i.e. the scaling length of the process.
However in superdiffusive processes, for $\alpha<1$, $f_{\alpha}(x)$ is a non trivial function 
decaying at large distances as $x^{-1-\alpha}$, so an analogous property is not trivial.

\begin{figure}
\includegraphics[width=\columnwidth]{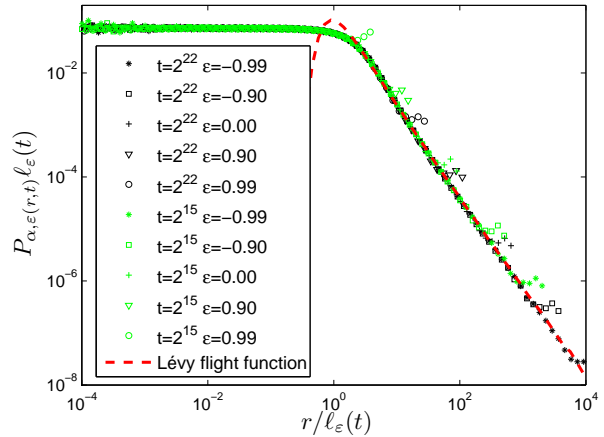}
\caption{Rescaling of $P_{\alpha,\varepsilon}(r,t)$ for $\alpha=0.7$ on a random structure. 
The value of $\ell_\varepsilon(t)$ are 
evaluated according to \eqref{ellcom} and the values of the coefficients $A_{\varepsilon}$ is given by \eqref{Aepsilon}. Red-dashed line represents the shape of the scaling function characterizing an uncorrelated L\'evy flight with the same $\alpha=0.7$. Notice that the tail of the distribution is the same since in both processes it is determined by  the value of $\alpha$.}
\label{scal0_7}
\end{figure}

Numerical data, in Fig. \ref{scal0_7} evidence for $\alpha=0.7$ in the random case that 
$P_{\alpha,\varepsilon}(r,t)$ for different values of the transmittance
can be scaled into a single function, independent of $\varepsilon$. 
Moreover, the dashed line 
shows that the scaling function is different from the standard L\'evy function 
describing the sum of independent L\'evy distributed random variables,  even if they are
characterized by the same long tail \cite{ann}.

\begin{figure}
\includegraphics[width=\columnwidth]{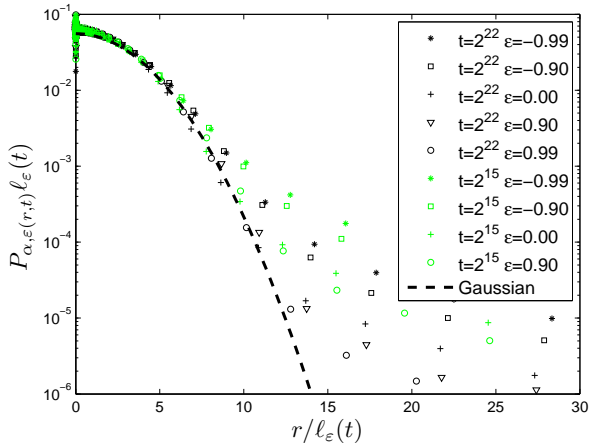}
\caption{Rescaling of $P_{\alpha,\varepsilon}(r,t)$ for $\alpha=1.5$ on a random structure. 
The value of $\ell_\varepsilon(t)$ are 
evaluated according to \eqref{ellcom} and the values of the coefficients $A_{\varepsilon}$ is given by \eqref{Aepsilon}. Dashed line represents the gaussian behavior of the scaling function for not too large $r/\ell_\varepsilon(t)$. At larger $r/\ell_\varepsilon(t)$ there is a slowly decaying function $h_{\alpha,\varepsilon}(r,t)$ superimposed to the gaussian behavior. At long times (black symbols) $h_{\alpha,\varepsilon}(r,t)$ tends to vanishes, however it still influences the momenta of the distribution. }
\label{scal1_5}
\end{figure}

\begin{figure}
\includegraphics[width=\columnwidth]{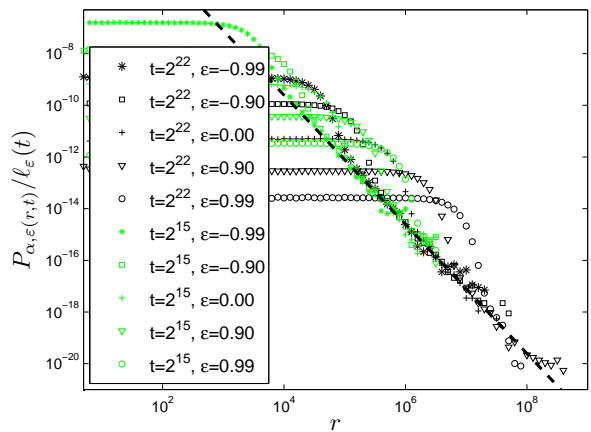}
\caption{Plot of $P_{\alpha,\varepsilon}(r,t)/\ell_\varepsilon(t)$ as a function of $r$ for $\alpha=1.5$ on a random structure. 
The value of $\ell_\varepsilon(t)$ are evaluated according to Equation \eqref{ellcom}. 
The plot evidences that the tail $h_{\alpha,\varepsilon}(r,t)$ of the distribution 
corresponds to Equation \eqref{hfunc} where the
effect of different transmission coefficient can be summarized in the constant 
$A_{\varepsilon}$ in \eqref{Aepsilon}.}
\label{scal1_5long}
\end{figure}

Let us now consider  the random case with $\alpha>1$. The scaling length grows as in a diffusive process,
and therefore we expect the scaling function $f_{\alpha}(x)$ to be a
Gaussian, independently of $\varepsilon$ and $\alpha$. 
Fig. \ref{scal1_5}, obtained for $\alpha=1.5$, evidences this property 
for small $r/\ell_\varepsilon(t)$. However,  for large $r/\ell_\varepsilon(t)$, 
$P_{\alpha,\varepsilon}(r,t)$ contains now the subleading contribution $h_{\alpha,\varepsilon}(r,t)$. This term
vanishes for large times but it can influence the momenta of the distribution.  
Fig. \ref{scal1_5} shows, at 
large $r/\ell_\varepsilon(t)$, 
the presence of this subleading term.
In \cite{Levyrandom},  $h_{\alpha,0}(r,t)$ has been evaluated using a "single long jump" 
approximation.  Here, we show that the same argument applies within the ansatz \eqref{ellcom} with the coefficients determined by \eqref{Aepsilon}, and
leads to a correct estimate of the function $h_{\alpha,\varepsilon}(r,t)$. In particular, within the same approximation,  for $r/\ell_\varepsilon(t)\gg 1$ the probability of reaching a point  at distance $r$ is
determined by the probability of performing a single ballistic stretch of length $r$, times the number of scatterers visited by the walker in a time 
$t$; this number can be estimated as $\ell_\varepsilon(t)/\Delta$, where $\Delta$ is the average 
distance between the scatterers, that for $\alpha>1$ is finite and independent of $\varepsilon$. 
If the whole effect of a variation in the transmittance $\varepsilon$ can be 
encoded in a change of  $A_{\varepsilon}$ as in Eq. \eqref{ellcom}, then we expect that 
the behavior of $h_{\alpha,\varepsilon}(r,t)$ can be estimated as 
\begin{equation}
h_{\alpha,\varepsilon}(r,t)\sim \frac{\ell_\varepsilon(t)}{r^{1+\alpha}}
\label{hfunc}
\end{equation}
where the proportionality constant is independent of $\varepsilon$.
In Fig. \ref{scal1_5long} we check indeed that our ansatz is correct and  that the tail of the distribution
$P_{\alpha,\varepsilon}(r,t)$ can be described according to equation \eqref{hfunc}, with the $\varepsilon$-dependent coefficient.

Finally, we focus on the case of deterministic one-dimensional fractal quasi-lattices i.e. the
lower panel of Fig. \ref{model}, where the step length distribution is described by the 
parameters $n_u$ and $n_r$.  As explained in \cite{Cantor1},
for $\varepsilon =0$ the motion of the random walker is ruled by the parameter
$\alpha=\log(n_u)/\log(n_r)$ which plays the role of $\alpha$ in the random structure. 
In particular,  the scaling length of the process 
also in the deterministic case grows as $t^{1/(1+\alpha)}$ for $\alpha<1$ and $t^{1/2}$
for $\alpha>1$. Here, we consider averages performed 
over processes starting from a given point of the structure, e.g the origin $0$ evidenced 
in Fig. \ref{model}. For local quantities, the scaling function 
does not present long tails since arbitrary long jumps are placed far away from the starting
point. On the other hand, for $\alpha<1$ the fractality of the structure  induces characteristic 
log-periodic oscillations 
in the scaling function. In particular for $\varepsilon=0$ we have \cite{Cantor1}
\begin{equation}
P_{\alpha,0}(r,t)={\ell}_0^{-1}(t)f'_{\alpha,0}(r/\ell_0(t),g(\log_{n_u}\ell_0(t))) 
\label{sal_c}
\end{equation}
where $\ell_0(t)$ is the scaling length of the process on the quasi-lattice, $f'_{\alpha,0}$ is the scaling function and $g(x)$ is a function of period one. According to Eq. \eqref{ellcom}, a variation of the transmittance 
only induces a rescaling of the correlation length, so that
the scaling function for a generic $\varepsilon$ is expected to be :
\begin{equation}
P_{\alpha,\varepsilon}(r,t)={\ell}_\varepsilon^{-1}(t)f'_\alpha(r/\ell_\varepsilon(t),g(\log_{n_u}\ell_\varepsilon(t))) 
\label{sal_cant_ep}
\end{equation}
where $\ell_\varepsilon(t)$ and the corresponding coefficients are again given by equations (\ref{ellcom},\ref{Aepsilon}).

Therefore for processes with different transmittances $\varepsilon$ and $\tilde \varepsilon$,
scaling holds if times $t$ and $\tilde t$ are chosen so that 
$\log_{n_u}\ell_\varepsilon(t)=k+\log_{n_u}\ell_{\tilde{\varepsilon}}(\tilde{t})$ with $k$ integer,
i.e. 
\begin{equation}
\tilde{t} =\frac{(1+\varepsilon)(1-\tilde{\varepsilon})}{(1-\varepsilon)(1+\tilde{\varepsilon})}(n_un_r)^k.
\label{times}
\end{equation}

\begin{figure}
\includegraphics[width=\columnwidth]{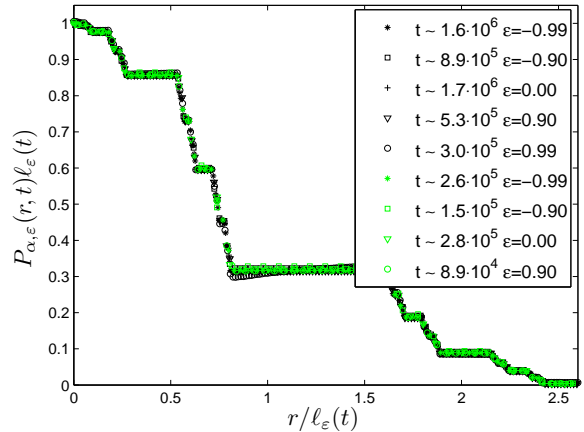}
\caption{Rescaling of $P_{\alpha,\varepsilon}(r,t)$ for $\alpha=\log(2)/\log(3)$ on a fractal quasi-lattice. 
The value of $\ell_\varepsilon(t)$ are 
evaluated according to (\ref{ellcom},\ref{Aepsilon}). Times are chosen
according equation \eqref{times} to detect scaling.}
\label{scallcantor}
\end{figure}

In Fig. \ref{scallcantor} we evidence that for times chosen according to equation \eqref{times}
scaling holds for different  $\varepsilon$. The complex devil staircase shape of 
the scaling function is typical of these fractal structures. Notice that the choice of times \eqref{times}, as a function of $\varepsilon$,  is crucial to recover the scaling, a further test of the validity of Eq. \eqref{ellcom}. In the case $\alpha>1$ the scaling functions are Gaussians, log-periodic oscillation are absent and scaling can be recovered through \eqref{ellcom} and \eqref{Aepsilon}, without the tuning of times \eqref{times}, as in  the random case.

\section{\label{sec:timeres} Time resolved transmission}

Due to its generality, the scaling approach can be useful not only in the analysis of
$P_{\alpha,\varepsilon}(r,t)$ but also for other interesting physical quantities, such as the exit
time probability and the time resolved intensity \cite{buonsante}, which are experimentally more
relevant.  Let us consider the effect of a transmittance coefficient $\varepsilon\not=0$ in the
case of a random structure. We consider a walker starting at time $t=0$ from the border of a sample of
size $L$ and we define the time resolved transmitted intensity $I_{\alpha,\varepsilon}(t,L)$ as the probability for the
walker to reach the boundary at distance $L$ before returning to the starting point.
The scaling hypothesis for the intensity $I_{\alpha,\varepsilon}(t,L)$ reads $I_{\alpha,\varepsilon}(t,L)=B g (t A_{\varepsilon}^{1+\alpha}/L^{1+\alpha})$ where $g(\cdot)$  is a scaling function and the proportionality constant $B$ depends on $L$, $\alpha$ and $\varepsilon$.
We calculate the coefficient $B$ as follows.

\begin{figure}
\includegraphics[width=\columnwidth]{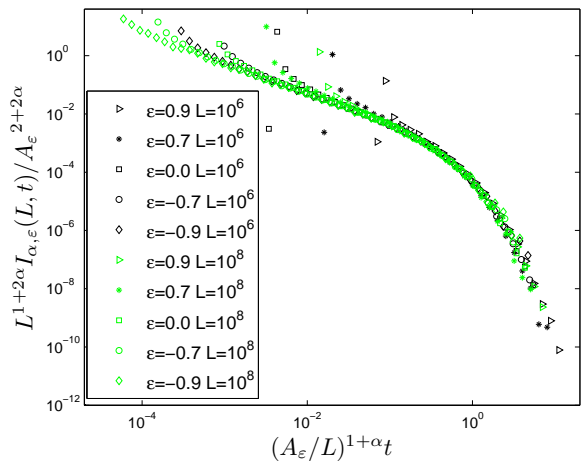}
\caption{The rescaled time-resolved transmitted intensity as a function of the rescaled time according equation \eqref{tri}, on a random structure with $\alpha=0.4$.}
\label{transm}
\end{figure}

The conductivity $C_{\alpha,\varepsilon}(L)$ is by definition the total number of walkers escaping from a system of size $L$ before returning to the starting point, independently of time i.e.
\begin{equation}
C_{\alpha,\varepsilon}(L)=\int_0^\infty I_{\alpha,\varepsilon}(t,L) dt.
\label{cond}
\end{equation}
Imposing the integral \eqref{cond} to be proportional to $A_{\varepsilon}^{1+\alpha}/L^{\alpha}$ according to equation \eqref{cond3}, we obtain
\begin{equation}
I_{\alpha,\varepsilon}(t,L)= L^{-2\alpha-1} A_{\varepsilon}^{2+2 \alpha} g(t A_{\varepsilon}^{1+\alpha}/L^{1+\alpha}).
\label{tri}
\end{equation}
The scaling form \eqref{tri} has been verified in Figure \ref{transm} for $\alpha=0.4$ and different values of $\varepsilon$, with the values of the coefficients determined by \eqref{Aepsilon}.

\section{\label{sec:cocl} Conclusions}

We have shown that L\'evy walk in a quenched, long-range correlated structures satisfies
a generalized scaling relation for arbitrary values of  the transmittance, both in the random and in the
fractal case, as expressed by Eq. (\ref{sal}) and (\ref{sal_cant_ep}) respectively. The main difference
between the two models is in the presence of the subleading term for
the random case and of log-periodic oscillations in fractal quasi-lattices.  As expected, all the leading
scaling behavior  are unaffected by a change of the scattering length  (the parameter
$\varepsilon$ in our model). This parameter enters in the multiplicative prefactor, 
ruling the dependence of the scaling
length as a function of time. Estimation of the latter is  of course relevant for finite samples and times.

We obtained an analytic expression for the multiplicative coefficients of the scaling lengths,
as a function of $\varepsilon$ in the annealed  L\'evy walk case, evidencing its independence on the transmittance in the superdiffusive regimes. Within the scaling framework, we also determined a closed form for the coefficient in the quenched random and fractal cases, which are the most relevant for experiments.

Another remarkable result of our analysis is that the scaling functions feature a superscaling property, 
namely they are independent of the transmittance and only depend on the structure through the
exponent $\alpha$. Interestingly, we employed the scaling properties to infer the dependence of
the time-resolved transmission in finite samples. 
All our analytic results have been compared with numerical simulations, with excellent agreement.

As the scaling picture discussed here has been shown to hold also in higher dimensional cases \cite{buonsante}, we expect that this superuniversality,
holding in one dimension,  can be detected also in higher dimensions. Moreover, we expect that superuniversality could hold not only for a variation of the transmittance $\varepsilon$ but for a wider class of local transformation of the dynamics such as the introduction of waiting times or of second neighbors jumps.

Besides their theoretical interest, our result are of importance
to interpret correctly the experimental and numerical results.
For instance, in the optical experiments of Ref.~\cite{bertolotti} it
is possible to control the mean free path of light in
diffusive media and investigate the approach to the scaling 
limits for the same distribution of glass sphere diameters.

\begin{acknowledgments}
This work has been partially supported by the MIUR Project 
{\it P.R.I.N. 2008} ``Nonlinearity and disorder in classical and
quantum processes.''  and by the  MIUR Project 
{\it P.R.I.N. 2008}  project "Efficienza delle macchine termoelettriche: un approccio microscopico". The authors acknowledge useful discussions with P. Buonsante.
\end{acknowledgments}

\end{document}